# The Cross Correlation between the Gravitational Potential and the Large Scale Matter Distribution

**Søren Madsen[1], Andrei G. Doroshkevich[2], Stefan Gottlöber[3] and Volker Müller[3]**

[1] Copenhagen University, Astronomical Observatory, Juliane Maries Vej 30, DK-2100 Copenhagen Ø, Denmark
[2] Theoretical Astrophysics Center, Juliane Maries Vej 30, DK-2100 Copenhagen Ø, Denmark
[3] Astrophysikalisches Institut Potsdam, An der Sternwarte 16, D-14482 Potsdam, Germany

**Abstract.** The large scale gravitational potential distribution and its influence on the large–scale matter clustering is considered on the basis of six simulations. It is found that the mean separation between zero levels of the potential along random straight lines coincides with the theoretical expectations, but it scatters largely. A strong link of the initial potential and the structure evolution is shown. It is found that the under–dense and over–dense regions correlate with regions of positive and negative gravitational potential at large redshifts. The over–dense regions arise due to a slow matter flow into the negative potential regions, where more pronounced non–linear structures appear. Such regions are related to the formation of huge super-large scale structures seen in the galaxy distribution.

## 1. Introduction

The tremendous growth of observational data during the last few years has strongly changed our insight into the structure and evolution of the universe. The large scale matter distribution of the universe has been predicted by the nonlinear theory of gravitational instability of Zeldovich (1970), and the Large Scale Structure (LSS) has already been identified in the first wedge diagrams of galaxy redshift surveys (Gregory & Thompson 1978) underlining qualitatively the theoretical picture.

During the last decade such phenomena as the Great Void (Kirshner et al. 1983), the Great Attractor (Dressler et al. 1987) and the Great Wall (de Lapparent et al. 1988) were extraordinary examples of the hugest observed structures in the Universe. However, the analysis of the presently largest galaxy catalogue – the Las Campanas Redshift Survey (Shectman et al. 1996) – demonstrates clearly that distinguished very large structure elements accumulate about 50% of all galaxies of the survey (Doroshkevich et al. 1996, 1997b). These structure elements are in many respects similar to the Great Wall and the superclusters of galaxies, we call them super-large scale structure (SLSS). The existence of these structures needs to be studied theoretically and in simulations in order to understand and to explain an essential feature of the universe.

The large scales (about 100 $h^{-1}$ Mpc) by which the SLSS is characterized make this problem especially enigmatic, since the galaxy correlation length is smaller than 10 $h^{-1}$ Mpc, and the cluster correlation length is smaller than 25 $h^{-1}$ Mpc. The matter distribution is, according to the general conviction, approximately homogeneous at larger scales. This point of view now needs to be revised. Perhaps, however, the galaxy concentration within the SLSS elements is not accompanied by a similar concentration of dark matter (see discussion in Piran et al. 1993, Demianski & Doroshkevich 1997), i.e. a large scale bias between dark matter and galaxies could exist.

The formation of the SLSS elements can be explained by matter infall into large scale gravitational wells. Undoubtedly, this explanation is valid for the description of the formation of a single wall and, hence, the observed galaxy distribution maps the well distribution. Therefore, the main problem in question is the origin of such wells, and the explanation of their spatial distribution. This aspect becomes especially important in view of new huge redshift surveys under preparation.

Apparently potential, velocity and density provide us with formal equivalent descriptions of the matter evolution because they are related by well known equations. For example, the adhesion approach, which operates only with the initial potential, describes very detailed the matter evolution (see, e.g., Shandarin & Zeldovich 1989). On the other hand, in a seminal paper, Bardeen et al. (1986) consider the same problem using the density field.

However, all previous experience in physics shows, that seemingly equivalent descriptions emphasize different aspects of the same problem and, thus, they are supplementary to each other. In practice the density field is more representative for the small scale evolution while the potential is more suited for the investigations of large scale matter evolution. Both descriptions need to be used together to reveal the interaction and the mutual influence of long and short wave perturbations and to obtain the general description of the matter evolution.

It is important that due to very slow evolution of the large scale spatial distribution even the initial potential field can be used for the prediction and explanation of properties of the matter distribution at later evolutionary stages. In particular, such analysis allows us to specify the large scale perturbations responsible for the SLSS formation (Demianski & Doroshkevich 1997).

A short theoretical analysis of the possible impact of the potential perturbations on the structure evolution has been given by Buryak et al. (1992) and by Demianski & Doroshkevich (1992). There a correlation of potential and density perturbations during all evolutionary periods was demonstrated quantitatively, and the typical scale of the potential perturbations was derived. Doroshkevich et al. (1997a) have improved some of the relations describing the potential perturbations. On the basis of six numerical simulation we will demonstrate here the close links of the spatial distribution of matter density with the potential distribution. Further we propose some quantitative characteristics of this correlation.

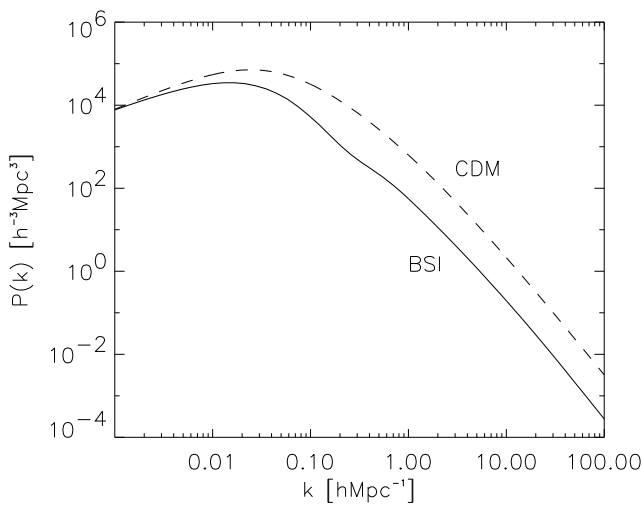

**Fig. 1.** The compound power spectra for CDM & BSI models

This paper is organized as follows. In Sect. 2 we describe the numerical models utilized for further analysis. In Sect. 3 the simulated potential distributions are discussed and characterized quantitatively. In Sect. 4 the link of potential and density perturbations is considered and some quantitative characteristics of this interactions are found. In Sect. 5 we summarize and discuss our results.

## 2. Numerical models

In this paper the spatial distribution and evolution of the potential is investigated using six numerical simulations of the standard CDM model (SCDM) and the double inflation model leading to power spectra with broken scale invariance (BSI) (Gottlöber et al. 1991). We have chosen these two models because they base both on cold dark matter, however they possess different power spectra on the relevant scales. Nevertheless, we found quite similar results for both models. Thus we present in the following mainly figures obtained from SCDM simulations, and we provide results both from CDM and BSI in the text and in the tables. We employ particle mesh simulations with $128^3$ particles in $256^3$ cells using three computational boxes with box sizes $L_{box} = 200$, $75$ & $25 h^{-1}$ Mpc (they are denoted by CDM200, CDM75 & CDM25 and BSI200, BSI75 & BSI25, respectively). A more complete description of the simulations used here is given in Kates et al. (1995). All simulations were started at redshift $z = 25$ using the same initial random phases, where the amplitudes were calculated in accordance with the first year COBE normalisation. The power spectra for CDM and BSI models are plotted in Fig. 1.

Let us note here that due to the finite resolution and finite box sizes all simulations deal with truncated power spectra. Speaking about the potential or the density in a PM-simulation, we always take the CIC values, i.e. the fields smoothed over the cell size. As measured with respect to the variance on the grid, the amplitude of initial perturbations is the largest for CDM25 and the smallest for BSI200. The simulations BSI75, BSI25, CDM200 and CDM75 are characterized by a successive growth of the initial amplitude between the extreme cases. These differences lead to a different degree of clustering at the same redshifts. For BSI200 the structure formation began not till redshift $z \approx 2$, and at the redshift $z = 0$ we see only weakly developed structures (Doroshkevich et al. 1997a). However, in CDM25 the disruption of structures and the formation of massive clumps can already be observed at $z \approx 2$. Thus with this sample of models we can investigate the structure evolution over a wide range of evolutionary stages.

## 3. Spatial distribution of gravitational potential

As the first step of our consideration we need to reveal and to describe the properties of the simulated potential

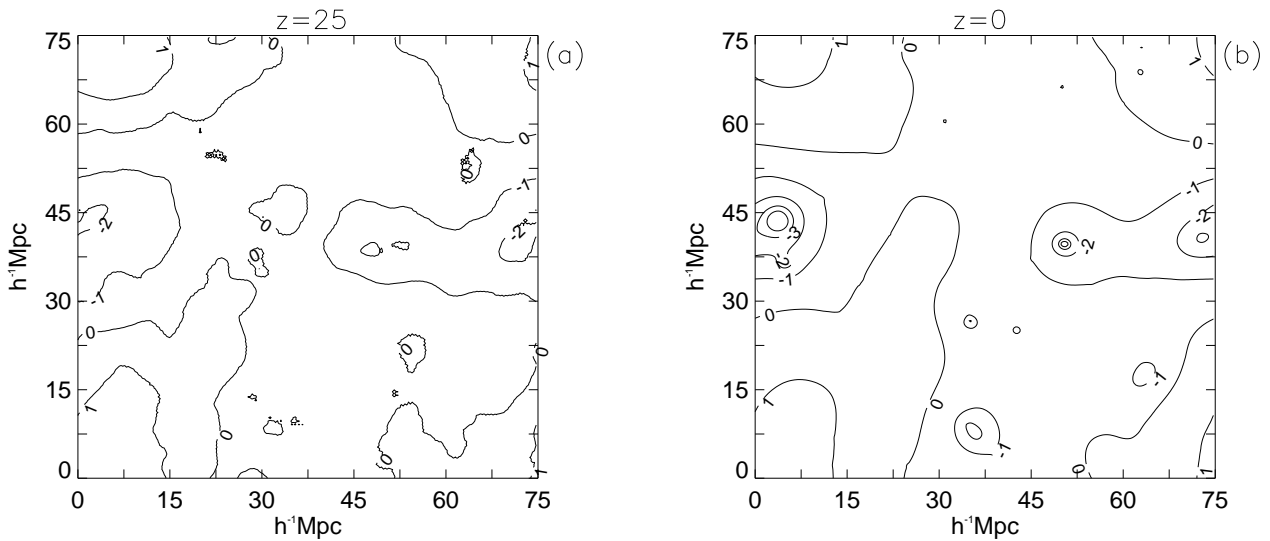

**Fig. 2.** Slices of the gravitational potential from CDM75 (left at $z = 25$, right at $z = 0$). The contour numbers are the values of the potential in units of $10^6$ (km/s)$^2$

distribution. We need also to introduce and to test some quantitative characteristics of the spatial distribution of the potential field.

The spatial distribution of random field can be conveniently characterized by various mean values that are combinations of moments of the power spectrum (some examples of such characteristics were given by Bardeen et al. 1986 and Sahni & Coles 1995). It is important to specify whether they are given for untruncated power spectra which is important for theoretical considerations and comparison with observations, or together with various window functions what is important for comparisons with simulations which operate with truncated power spectra.

### 3.1. Qualitative description

The main properties of the potential distribution can be demonstrated by the evolution of the spatial distribution of the potential in a slice plotted in Fig. 2 for CDM75 at $z = 25$ and at $z = 0$. The larger features of the potential are unchanged while on small scales the structure of potential is modified strongly and all the wrinkles in the equipotential lines are smoothed out with time. The lines come from the intersection of the plane under consideration with the 2-dimensional equipotential surfaces in 3-dimensional space.

The matter concentrations in denser clumps influences locally the potential. This can be clearly seen in the most negative regions of the potential in Fig. 2, where some very deep potential wells evolve with time. The deep hole in the centre of Fig. 2 (b) is especially prominent.

The isolines of potential distribution in Fig. 2 demonstrate also the well known fact, that the shapes of these isolines are more or less elliptical for high potential peaks and deep wells, and they become more and more complicated near the zero level. These properties are typical for random Gaussian fields which we now describe quantitatively.

### 3.2. Quantitative characteristics of the spatial potential distribution.

Let us consider the potential along a random straight line. Thus we transform the complicated three–dimensional problem to the much simpler one–dimensional one. A 1D potential distribution is plotted in Fig. 3 for CDM75 and CDM200. The CDM75 profiles in Fig. 3(a) correspond to the cuts $x \sim 22h^{-1}$Mpc of Fig. 2, so that the "blob" at $y \sim 55h^{-1}$Mpc of Fig. 2(a) is included. This feature is seen in Fig. 3(a) as a tangent to $\phi = 0$. Figure 3 demonstrates that in general, excluding the high dense clumps, the potential becomes smoother with time. Indeed, here the potential contains only one large wavelength giving raise to one maximum and one minimum. It is interesting that both the CDM75 and the CDM200 simulation contain just one basic wavelength of the potential perturbations which is a characteristic of the realized range of the CDM potential perturbations.

In the following we call regions with positive potential ($\phi > 0$) $\Phi^+$-regions. During cosmic evolution the density in these regions becomes lower than the mean density (under dense region - UDR). In $\Phi^-$-regions ($\phi < 0$) the density becomes higher than the mean density (over dense region - ODR). This can be clearly seen in Fig. 4 where the mean value of the density contrast in both regions is shown separately. Note, that the density contrast increases in $\Phi^-$-regions only by a factor of about 1.4 for CDM200. In the more evolved CDM25 simulation, we find over/underdensities of 2.2 and 0.17 for $\Phi^-$ and $\Phi^+$, respectively. More detailed properties of $\Phi^-$ and $\Phi^+$ regions are given later on (Table 2). There we will show that the density contrast arises from a small mass flow through the boundary between $\Phi^+$ and $\Phi^-$ regions, whereas the respective volumes change only by a very small amount.

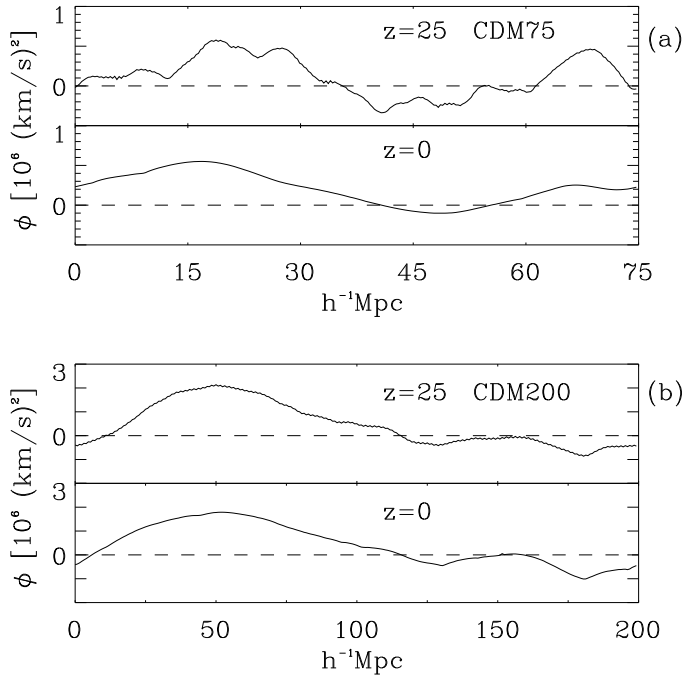

**Fig. 3.** One-dimensional potential distribution in CDM75 & CDM200

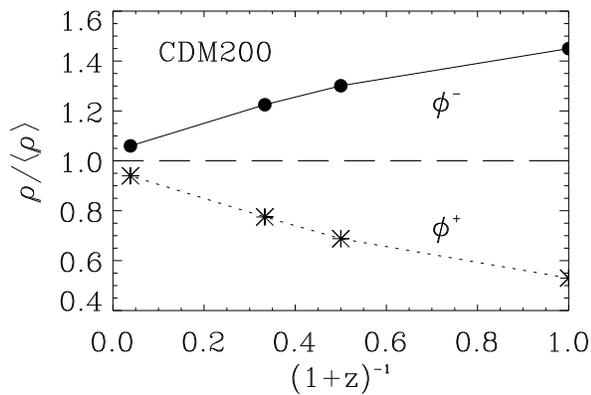

**Fig. 4.** Evolution of the density contrast in the regions of positive ($\Phi^+$) and negative ($\Phi^-$) potential for CDM200

The 1D approach allows us to introduce a simple quantitative characteristic of the potential distribution, the mean separation (mean free path) between zero points of the potential, $L_0$. Demianski & Doroshkevich (1992) found a relation between $L_0$ and the power spectrum $P(k)$,

$$L_0^2 = 3\pi^2 \frac{\int_0^\infty \frac{P(k)}{k^2} dk}{\int_0^\infty P(k) dk}, \tag{1}$$

where $k$ is the comoving wave number. The value $L_0$ is closely linked with the correlation length of the potential (see, e.g., Bardeen et al. 1986). Equation (1) can be used for simulations with truncated power spectra, however, it can obviously not be used for spectra with Harrison-Zeldovich asymptote $p \propto k$ at $k \to 0$, because the upper integral is divergent and leads to an infinite correlation length of the potential. Doroshkevich et al. (1997a) have derived a general relation which can be simplified for broad band power spectra to the equation

$$L_0^2 = 3\pi^2 \frac{\int_0^\infty \frac{P(k)}{k^2} \left(1 - \frac{\sin kL_0}{kL_0}\right) dk}{\int_0^\infty P(k) dk} \tag{2}$$

the solution of which gives the correct value $L_0$ both for the truncated and the total power spectrum. For the standard CDM and BSI matter dominated power spectra with $\Omega = 1$ we obtain

$$L_0^{CDM} = 31h^{-2}\mathrm{Mpc}, \quad L_0^{BSI} = 59h^{-2}\mathrm{Mpc}.$$

Note, that due to the scaling of the transfer function with $\Gamma = \Omega h$ the resulting $L_0$ of our models scales with $h^{-2}$.

The comparison of these results obtained for the theoretical spectra with similar estimates for simulations provides a simplest quantitative characteristic of the representativity of the simulations with respect to the large scale spatial potential distribution because both expressions (1) and (2) are sensitive to small $k$-values and to the size of the simulation box. Hence, it allows to estimate the impact of the computational box size on the matter evolution.

In simulations the mean separation of points $\phi = 0$ along a random straight line, $\langle L_0 \rangle$, can be found directly. Thus, if $L_i$ is the set of measured distances between levels $\phi = 0$ and $N$ is their total number, then the mean separation and dispersion $\langle \sigma_0 \rangle$ are given by

$$\langle L_0 \rangle = \frac{1}{N} \sum_{i=1}^N L_i \qquad \langle \sigma_0 \rangle = \sqrt{\frac{1}{N} \sum_{i=1}^N (L_i - \langle L_0 \rangle)^2}. \tag{3}$$

In Table 1 the results are listed for CDM and BSI simulations at redshift $z = 0$, $z = 2$ and $z = 25$ together with the theoretical values of $L_0$ calculated from Eq. (2) for the truncated spectra used for simulations. The upper

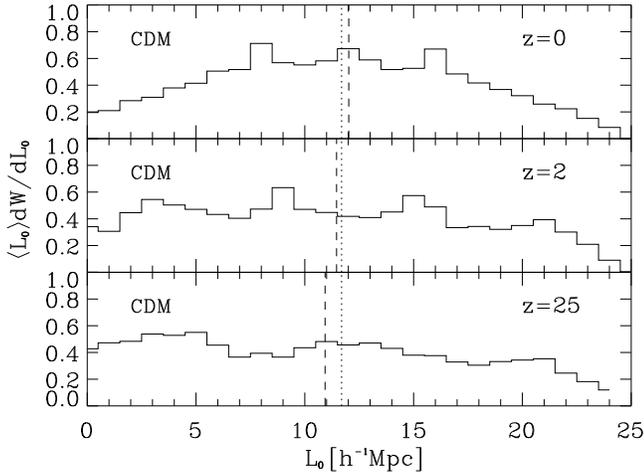

**Fig. 5.** Frequency distribution $\langle L_0 \rangle dW/dL_0$ of measured distances $L_0$ between potential levels $\phi = 0$ in the 25 $h^{-1}$ Mpc simulations. The dashed lines are the mean $\langle L_0 \rangle$ from Table 1, and the dotted lines are the theoretical values $L_0$.

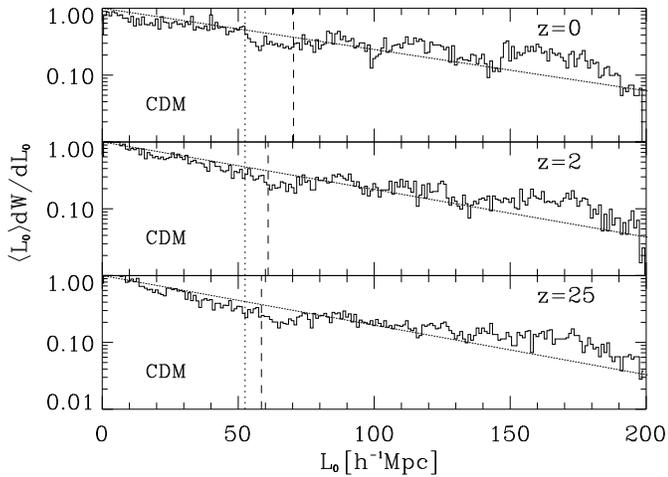

**Fig. 6.** Frequency distribution $\langle L_0 \rangle dW/dL_0$ of measured distances $L_0$ between potential levels $\phi = 0$ in the 200 $h^{-1}$ Mpc simulations. The dashed lines are the mean $\langle L_0 \rangle$ from Table 1, and the dotted lines are the theoretical values $L_0$. Fits to the histogram with the Poisson distribution are included as dotted lines.

and lower limits of the integral were taken from the box size and the cell size.

The dispersion in Table 1 characterizes the actual distance distribution rather than the errors in the measurements of $\langle L_0 \rangle$. The reason for the large dispersion becomes clear from frequency distributions $\langle L_0 \rangle dW/dL_0$ of the measured $L_i$ values (see Figs. 5 and 6 for CDM25 and CDM200, respectively). They are produced by 3963 random straight lines cutting the actual contour levels of the potential in the simulation box.

In CDM25 and BSI25 simulations the different distances between levels $\phi = 0$ are almost equal abundant, but we see already a trend towards smaller distances. This becomes more evident in the bigger simulations. Actually, the histogram is well described by a Poisson distribution in case of the CDM200 and BSI200 simulations (see the dotted line in Fig. 6).

**Table 1.** Theoretical and measured values of the mean separation $L_0$ and their dispersion $\sigma_0$ for potential levels $\phi = 0$

| | z | $\langle L_0 \rangle$ ($h^{-1}$Mpc) | $\langle \sigma_0 \rangle$ ($h^{-1}$Mpc) | $L_0^{\text{theory}}$ ($h^{-1}$Mpc) |
|---|---|---|---|---|
| BSI25 | 0 | 11.55 | 6.51 | |
| | 2 | 11.15 | 6.61 | 12.41 |
| | 25 | 11.05 | 6.61 | |
| CDM25 | 0 | 12.02 | 5.82 | |
| | 2 | 11.46 | 6.55 | 11.69 |
| | 25 | 10.93 | 6.74 | |
| BSI75 | 0 | 33.94 | 19.85 | |
| | 2 | 33.22 | 20.06 | 36.19 |
| | 25 | 32.97 | 20.18 | |
| CDM75 | 0 | 33.43 | 19.88 | |
| | 2 | 30.65 | 20.85 | 28.31 |
| | 25 | 29.05 | 21.10 | |
| BSI200 | 0 | 77.96 | 55.01 | |
| | 2 | 76.16 | 55.02 | 74.38 |
| | 25 | 74.68 | 55.18 | |
| CDM200 | 0 | 70.34 | 53.95 | |
| | 2 | 61.01 | 52.80 | 52.53 |
| | 25 | 58.53 | 52.67 | |

From a theoretical point of view, we expect a Poisson distribution of SLSS elements (and also of $L_i$) along a random straight line for distances $L_i \geq \langle L_0 \rangle$ where any correlation becomes negligible (White 1979, Buryak et al. 1991, Borgani 1996). However, we see that for the small box sizes the measured distribution is far from Poissonian. For distances ($L_0 \geq 0.5 L_{box}$) this is caused by the limitations of the simulations regarding SLSS elements because the number of long wave harmonics in the perturbations are strongly limited. However, in case of smaller scales the potential distribution is correlated and these correlations depend on the power spectrum. The CDM200 model shows the best indications of a Poissonian distribution because it contains the most SLSS elements in all the simulations. This can be easily seen from the values of $\langle L_0 \rangle$ in CDM200, which shows an average of 3 to 4 intersections with levels $\phi = 0$ per box length, while on the other hand the 25 $h^{-1}$Mpc simulations have only about two.

In this respect the CDM25 and BSI25 models are similar to the scale free models with a power index between $n = -2$ and $n = -2.2$ (see Fig. 1). Only for the models CDM200 and BSI200 deviations from the power law becomes important, so that we have to deal with a real broad band power spectrum. As we could see, this is also reflected by the potential characteristics. Thus, both for $L_{box}=75h^{-1}$Mpc and $L_{box}=25h^{-1}$Mpc the values $\langle L_0 \rangle$ are close to half the box size, $0.5L_{box}$. In all simulations besides CDM200, the realized values $\langle L_0 \rangle$ are much smaller than the values $L_0^{CDM}$ obtained for the untruncated power spectra. This means that only the CDM simulation in the larger computational boxes realistically reproduces the potential distribution, and therefore it can expected to describe the matter evolution on the SLSS scale, comparable with the observed matter distribution. In contrast, any simulations in smaller boxes have more or less methodical character.

A common feature of all histograms (see also Table 1) is the slow increase of $\langle L_0 \rangle$ during evolution due to non-linear effects whereas $L_0^{\text{theory}}$ is defined by the initial power spectrum. However, $\langle L_0 \rangle$ remains near to $L_0^{\text{theory}}$ during the whole evolution. The slightly faster evolution of CDM75 and CDM25 should be ascribed to the late evolutionary stage of these simulations. Indeed, some structure elements have been destroyed at $z = 0$, and the matter distribution starts to transform into a system of isolated clumps (see also the discussion in Doroshkevich et al. 1997a).

### 3.3. Potential–potential correlation.

The simple but important quantitative characteristic of the potential distribution is the two–point autocorrelation function. It allows us to obtain the characteristics of potential distribution averaged over the entire simulation.

For the primordial Harrison–Zeldovich spectrum the 'gravitational potential' cannot be used directly because the dispersion of the potential diverges in the limit $k \rightarrow 0$. Therefore, here one has to use the 'potential difference' between two points. This implies some modifications of the theoretical description (see, e.g., Demianski & Doroshkevich 1997). However, in case of simulations this problem disappears due to the finite box size. Thus, we can use here the 'potential' without any restriction.

Thus, the standard correlation function of the potential in points $q_1$ and $q_2$ can be written as following:

$$\xi_\phi(z,q) = \frac{1}{(2\pi)^3} \int \frac{P(z,k)}{k^4} e^{i\mathbf{k}\mathbf{q}} d^3k$$
$$= \frac{1}{2\pi^2} \int_0^\infty \frac{P(z,k)}{k^2} \left( \frac{\sin kq}{kq} \right) dk, \qquad (4)$$

where $q = |\mathbf{q}_1 - \mathbf{q}_2|$. For CDM-like power spectra this relation can be fitted as following (Demianski & Doroshkevich 1997)

$$\xi(z,q) = \sigma_\phi^2(z) \left( 1 - \ln(1 + 3.6(q/L_0)^2) \right) \qquad (5)$$

where the value $L_0$ is the mean separation of the zero potential along a random straight line defined by Eq. (2) and

$$\sigma_\phi^2(z) = \frac{1}{(2\pi)^3} \int \frac{P(z,k)}{k^4} d^3k \qquad (6)$$

is the dispersion of the potential perturbations.

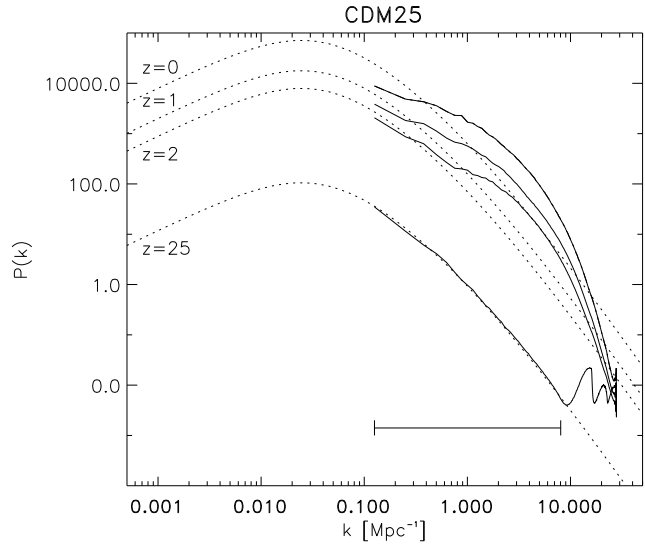

**Fig. 7.** The evolution of the power spectrum with redshift for CDM25. The dashed lines are the corresponding power spectra calculated by linear theory. The horizontal bar indicates box size and Nyquist frequency.

In Fig. 7 the evolution of the power spectrum is depicted for CDM25 at redshifts $z = 25, 2, 1$ & $0$. At high redshifts the power spectrum has some unphysical fluctuations at short wavelengths, caused by the finite resolution of the simulation. As upper limit in the integration of the power spectrum, we take the Nyquist frequency. Later the evolution increases strongly the short wave amplitude relatively to the linear law. In our case the long wave spectral amplitude decreases slightly due to the special initial realization.

The function $\xi_\phi(z,q)/\sigma_\phi^2(z)$ has been calculated for the CDM200 and CDM25 models using the power spectra reconstructed at several redshifts ($z = 25, 2$ & $0$) from the simulation. The results are presented in Fig. 8 together with the theoretical values given by Eq. (5). The different behaviour of theoretical and measured values at large distances characterizes the different potential distribution for truncated power spectra realized in simulations and the theoretical CDM spectra, i.e. it is caused by the impact of the finite box size. This conclusion is consistent to the former discussion of the one-dimendional analysis of separations of zero-points of the potential since this sep-aration concerns the length scales larger than half of the simulation box.

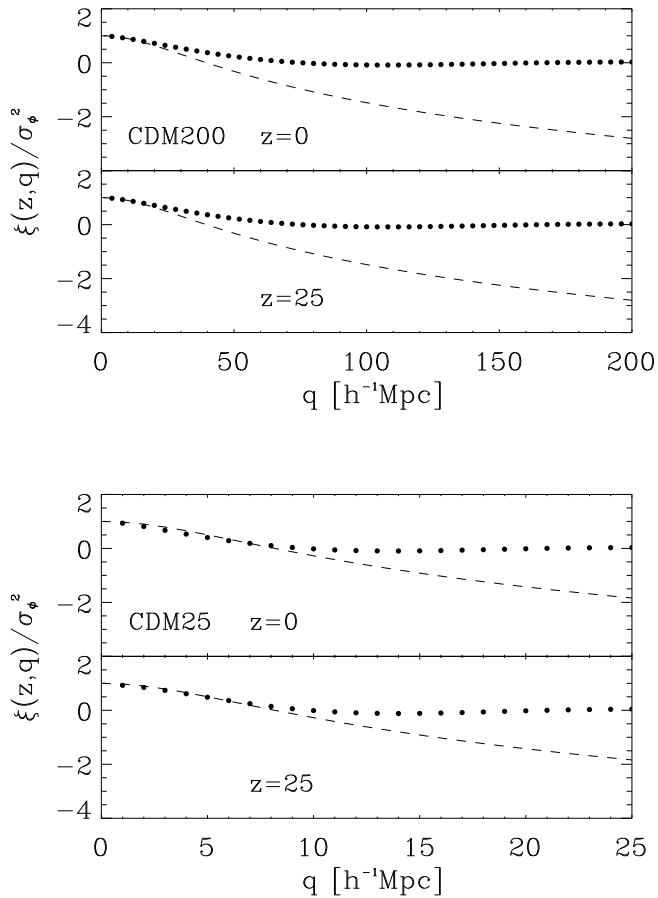

**Fig. 8.** The normalized correlation function of potential vs. comoving distance for two models and two redshifts. The dotted lines present the simulation results and the dashed line the expectation of Eq. (5).

## 4. Comparison of the spatial distribution of density and potential

The theory of gravitational instability as the origin of large scale structure is based on the close connection of the gravitational potential and the matter density field. However, this general statement has to be refined to discriminate between the influence of the potential on the density evolution and vice versa. Very roughly, we expect a leading role of density perturbations on small scales, and a leading role of potential perturbations on large scales. For example, the potential perturbations are responsible for the matter distribution in richer filaments (LSS) and in wall like structures (SLSS). As a rule, these structures map the potential wells. On the other hand, the thickness of these structure elements is due to dynamical effects in the evolution of the density field, and a theoretical model has to be used to trace it back to the characteristics of the potential field.

While the spatial distribution of the potential is driven by the large scale part of the power spectrum and, hence, evolves relatively slowly, the density distribution is mainly influenced by the small scale part of the power spectrum, and it evolves much faster. However, it can be shown that the distribution of the small scale density inhomogeneities is closely connected to the large scale spatial potential distribution.

This relation becomes evident considering the density field smoothed on a large scale. Then the evolution of both fields is controlled by the same part of the power spectrum. But the gravitational potential leads to a large scale modulation of the matter distribution, i.e. the much smoother potential provides a large scale imprint on the density field. Hence, it is natural to investigate this type of interaction avoiding any additional smoothing and to consider the density field given on the simulation grid in comparison with the potential. In this manner we can test more clearly the emergence of a *natural bias* or of resulting large typical scales in the matter density, and possibly in the spatial galaxy distribution leading to the wall-like SLSS. Here, we want to describe this link qualitatively and quantitatively using numerical simulations, a more detailes theoretical description should be given elsewhere.

### 4.1. Qualitative description

Contrary to the large scale potential field, the density perturbations evolve strongly both in the linear and the non-linear regime, cp. Fig. 9 for one simulation (for the same where the potential slice is shown in Fig. 2). We can clearly see an evolution towards more pronounced structures. This evolution looks like strings of clumped matter points moving toward the biggest clumps (clusters). These big clumps are pulling the smaller clumps along the inward pointing strings or filaments. Further we see richer filaments at redshift $z = 2$ than at $z = 0$. They seem to become less well defined with time, while at the same time large areas are more or less empty of matter, so that small voids appear.

In Fig. 10 we show the high density regions of the same CDM75 slice as in Fig. 9 at $z = 0$ for different contour levels. The slice corresponds to the gravitational potential field shown in Fig. 2(b). The contour levels include 50% of the total matter in the slice in (a), 25% in (b) and 10% in (c). While the contour levels of the dark matter shown in 9(c) have a clear filamentary structure, they disappear gradually if looking at the high density regions (Figs. 10(a) and (b)). Figure 10(c) contains only a few dense clumps, which can be interpreted as clusters, or at least as dark matter halos. From Fig. 10 we conclude that at $z = 0$ more than 25% of all matter in this slice is concentrated in a relatively low number of high density peaks in accordance with the estimates of Doroshkevich et al. (1997a). This is a typical feature of the considered late evolutionary stage. From Fig. 10 we also conclude that the spatially biggest clusters are also the heaviest.

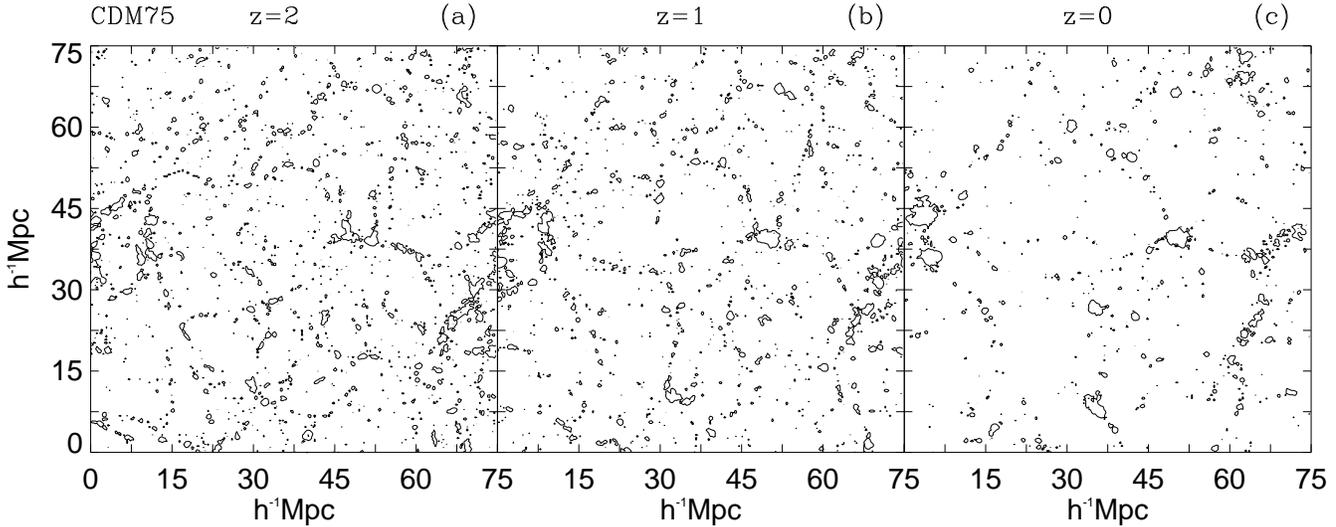

**Fig. 9.** The evolution of the density field with time from $z = 2$ to $z = 0$ in comoving coordinates. The contour levels are the same for all three redshifts

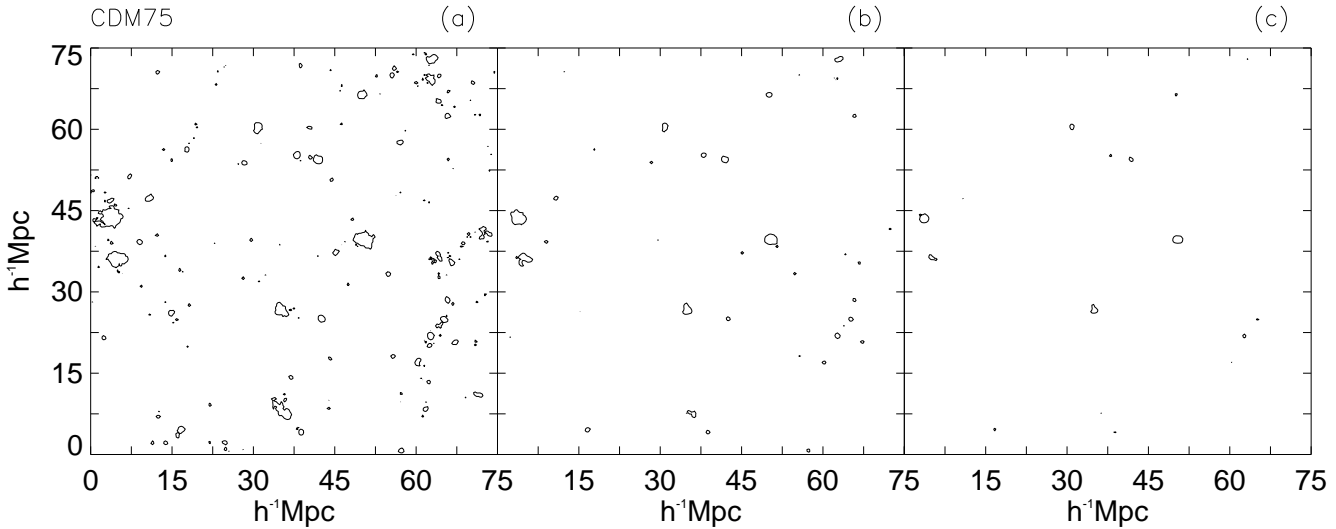

**Fig. 10.** A density slice for the CDM75 simulation, taken at $z = 0$. It is the same slice as in Fig. 9 but there the high density regions at $z = 0$ correspond to a fraction 75% of the total mass in the slice, here we show 50% in (a), 25% in (b), and 10% in (c).

Let us now compare Figs. 2, 9 and 10. In Fig. 9 we see the motion of smaller clumps and their merging with bigger clumps to the double cluster at the left, another cluster in the bottom and a third one near the middle. These big clusters have their counterparts in Fig. 2(b), where 3 very deep potential wells confine the clusters formed by $z = 0$. It is worth noting again that the evolution of the density field changes mostly the local potential distribution, but there are no global changes of the $\phi = 0$ contour levels with time. Also not much matter is flowing across these levels.

Comparing the gravitational potential in Fig. 2(b) with the contours of the density field in Fig. 10, we see that the weakest objects disappear with increasing contour levels, and, as expected, only the heaviest remain in the deepest potential wells. The highest density peaks are clearly correlated with the ODR. Thus the largest clusters trace the ODR.

The strong link between the spatial distribution of the density perturbations and the potential field is illustrated in Fig. 11, where a concentration of density peaks in regions of negative potential becomes obvious.

From the histograms in Figs. 5 and 6 we concluded that from all our simulations CDM200 had the greatest number of intersections with potential levels $\phi = 0$ along a random straight line, and we should therefore expect to find the highest number of SLSS elements in this model. In order to get a visual impression of the SLSS in the simulations, 3 redshift realizations for the two $200h^{-1}$ Mpc models have been plotted (Fig. 11). The first obvious property of the sequence of slices is the static nature of the levels $\phi = 0$ (the solid lines), they practically do not change their positions in space from $z = 2$ to $z = 0$, and so they are unaffected by the large-scale motion of matter. We also see that the evolution of the density field with time in $\Phi^-$-regions is faster than in $\Phi^+$-regions.

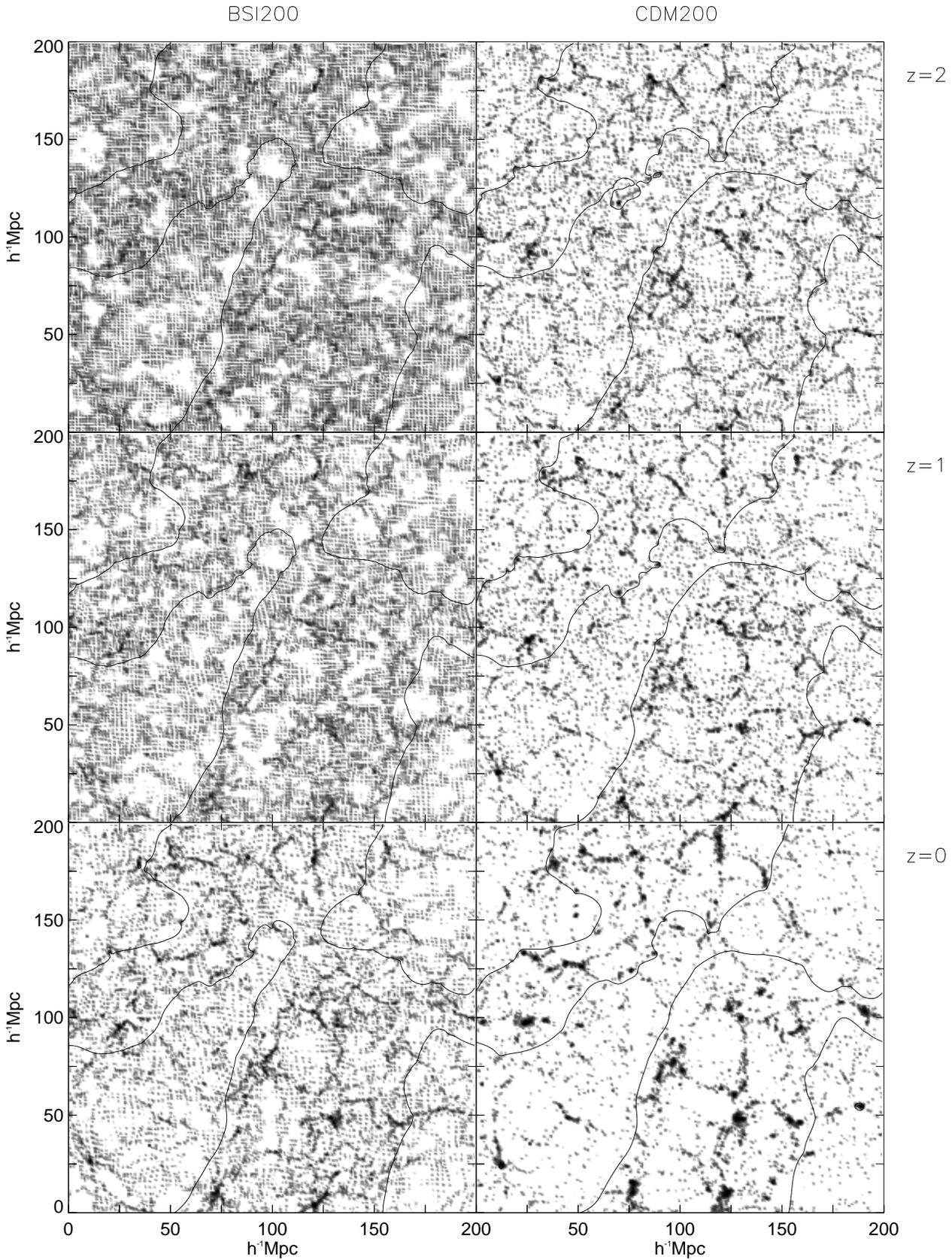

**Fig. 11.** The evolution of a slice from $z = 2$ to $z = 0$ of the BSI200 and CDM200 models, respectively. The density is shown as grey tones, where white corresponds to empty space and the darkest spots to the highest density peaks. The solid lines correspond to $\phi = 0$. The UDR are situated in the corners. The ODR at $z = 0$ are clearly seen as it accumulates the main fraction of high dense clumps.

The difference between the two cosmological models is likewise evident. This is not surprising, as the CDM model has more power than the BSI model on all scales covered by the simulation (see Fig. 1). Indeed, we see from Fig. 11 that the CDM model at $z = 2$ behaves approximately like BSI at $z = 0$. Not only the density peaks grow faster in the CDM model, but also the LSS forms earlier. Moreover, the LSS elements seem to break up already at $z = 0$. In the BSI model all evolution is later and slower. On larger scales the two models become more alike, which is also expected from the behaviour of the power spectrum. The placement of the potential levels $\phi = 0$ in the two models are practically the same, because the same random realization was used at $z = 25$. The stronger evolution of the CDM model is also seen by the larger number of clusters than in the BSI model. In the CDM model well defined clusters are already present at $z = 2$, but they do just first appear at $z = 0$ in the BSI model.

It is interesting that one heavy cluster has developed in the $\Phi^+$-region (lower left corner in Fig. 11) that shows that several dark matter clumps may arise also in $\Phi^+$-regions. This is a natural consequence of the random character of the perturbations.

Using the periodicity of the simulation box, we can identify a connected (i.e. a percolating) structure which seems to lie approximately in the centre of the $\Phi^-$-region in the CDM model at $z = 0$. Especially the highest density peaks (the darkest spots) seem to trace this structure. As mentioned earlier, we would expect to see about 1 to 2 walls along a straight line, however, the evolution of the CDM model on small scales already breaks up this wall-like structures.

Voids of different sizes are present. The mean void size grows with time while the number of voids decreases. It looks as though there are two kind of voids in Fig. 11. One type is bounded by the weaker structure, and the other bounded by the strongest density peaks. The latter defines a kind of super-voids inhabited by relatively weak filamentary structures.

The comparison of Figs. 2, 9, 10 and 11 point to the growth of the correlation between matter density and the gravitational potential due to the matter concentration into high dense clumps. At the same time the strong difference in the evolutionary rate of potential and density perturbations can be clearly seen. During the period under consideration the potential perturbations are changed on small scales, while on larger scales the potential distribution is very stable.

## 4.2. Quantitative description of the potential–density interaction.

The detailed quantitative characteristics of the potential–density interaction requires a theoretical derivation based on some approximation of the non-linear dynamics. As a first step, we only provide a description and general characterisation of this interaction.

Figure 12 illustrates this interaction for the models CDM200. The other models were investigated as well and they show similar results. The functions $f_{vl}$ and $f_{ml}$ are defined as the cumulative fraction of volume and mass, respectively, lying below the potential level $\phi_l$ whereas $\rho_l$ is the mean density in this region. Figure 12(a) shows the $f_{vl}$ versus $f_{ml}$ relation. The lower value of the fraction of volume $f_{vl}$, occupying the regions below the potential level $\phi_l$, than the value of the corresponding fraction of mass $f_{ml}$ shows the formation of ODR and UDR, the same behaviour is seen in Fig. 4. The matter concentrates in negative potential regions during the evolution. When more and more matter is gathered, and it forms deep potential wells, then the curves in Fig. 12(a) turn towards the right low corner. In Fig. 12(b) we show that the density increases by 3 orders of magnitude basically in the negative potential region. This is accompanied by the decrease of the global minimum of the potential.

Figure 12(c) shows the evolution of the differential distribution of the matter and space fraction, $f_m$ and $f_v$, respectively, at potential value $\phi$ for two redshifts. At $z = 25$ both curves are symmetrical with respect to the zero point, while at $z = 0$ only $f_v$ remains almost symmetrical and $f_m$ becomes significantly asymmetric. The typical wings of $f_m$ into the $\Phi^-$ regions arise from the matter infall into the potential wells while the shift of the median of the curve describes the slow mass flow through the $\phi = 0$ level.

A most interesting problem is the connection of the initial potential distribution with the current density distribution as it manifests the influence of initial perturbations to the later evolutionary stages. Such an influence lies on the basis of the adhesion model (see, e.g., Shandarin & Zeldovich 1989). The correlation between the potential at $z = 25$ and the density field at redshifts $z = 25$, 2 and 0 is shown in Fig. 13 for CDM25 and CDM200. Again one notices the slow flow of matter with time to the regions of negative initial potential. As it was shown above qualitatively, faster and more effective clustering occurs in regions of negative potential while the boundaries $\phi = 0$ remain practically at rest. Figure 13 gives us the quantitative characteristic of this effect and shows that the matter flow across the surface $\phi = 0$ in CDM200 is small. On the other hand, a stronger clustering and strong matter concentration in ODR occurs in the smaller simulation CDM25 which is in the latest non-linear evolutionary period. A significant number of high peaks are present at $z = 0$. They appear when the sheetlike and filamentary DM structures disrupt into a system of massive clumps (Doroshkevich et. al 1997a). Nevertheless, the strong correlation of the matter distribution with the initial potential distribution can be seen also in the small box. As expected, the correlation increases during the evolution.

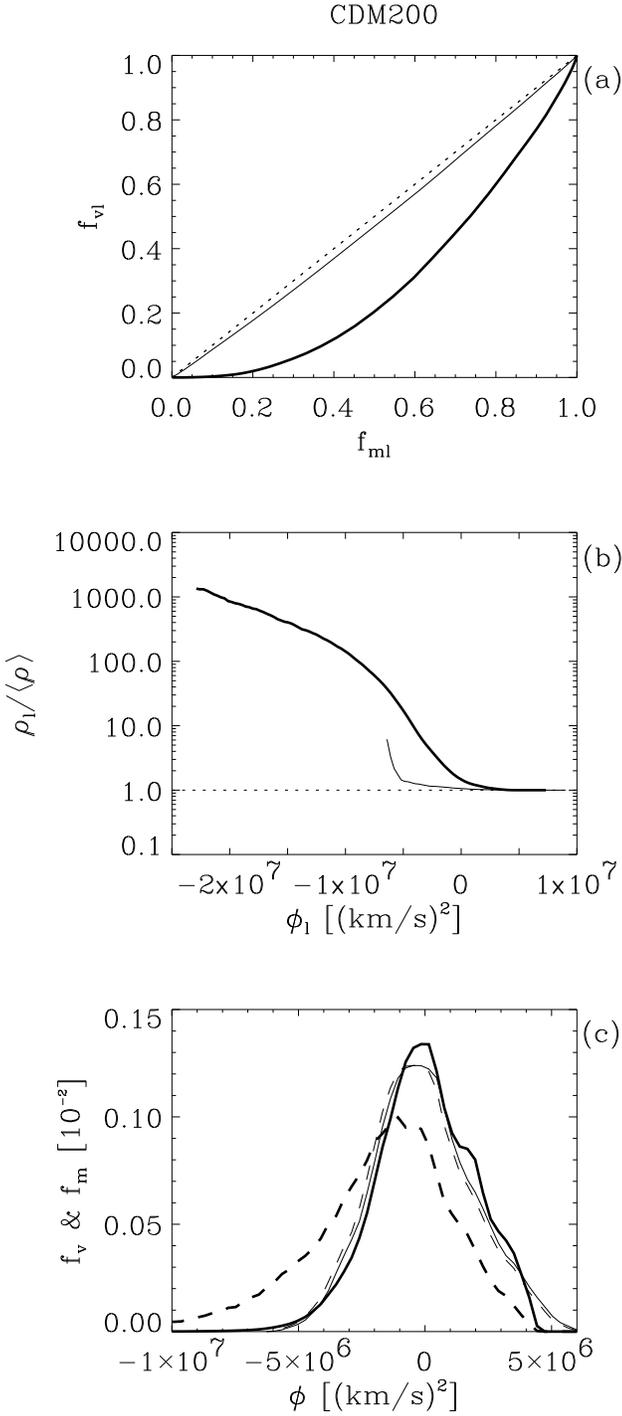

In Table 2, the fraction of mass $f_m$, the fraction of space $f_v$ of ODR and UDR, and the mean density $\rho$ in ODR and UDR, respectively, are shown for both the initial and final redshifts versus the initial and current potential distribution.

**Table 2.** Parameters of the ODR and UDR, defined as regions where $\phi < 0$ and $\phi > 0$, respectively for redshifts $z = 25$ and $z = 0$. The columns 25-0 denotes the comparison of the density field at $z = 0$ with respect to the initial potential field at $z = 25$ (for $f_v$ the columns 25 and 25-0 are identical). $f_m$ is the fraction of mass, $f_v$ is the fraction of space and $\rho$ is the density (in units of the critical density)

|  |  | ODR | | | UDR | | |
|---|---|---|---|---|---|---|---|
| z |  | 25 | 25-0 | 0 | 25 | 25-0 | 0 |
| BSI25 | $f_v$ | 0.54 | 0.54 | 0.48 | 0.46 | 0.46 | 0.52 |
|  | $f_m$ | 0.56 | 0.77 | 0.76 | 0.44 | 0.23 | 0.24 |
|  | $\rho$ | 1.04 | 1.43 | 1.58 | 0.96 | 0.50 | 0.46 |
| CDM25 | $f_v$ | 0.54 | 0.54 | 0.41 | 0.46 | 0.46 | 0.59 |
|  | $f_m$ | 0.58 | 0.91 | 0.90 | 0.42 | 0.09 | 0.10 |
|  | $\rho$ | 1.07 | 1.68 | 2.20 | 0.91 | 0.20 | 0.17 |
| BSI200 | $f_v$ | 0.53 | 0.53 | 0.52 | 0.47 | 0.47 | 0.48 |
|  | $f_m$ | 0.59 | 0.63 | 0.63 | 0.41 | 0.37 | 0.37 |
|  | $\rho$ | 1.11 | 1.18 | 1.21 | 0.87 | 0.79 | 0.77 |
| CDM200 | $f_v$ | 0.52 | 0.52 | 0.51 | 0.48 | 0.48 | 0.49 |
|  | $f_m$ | 0.55 | 0.70 | 0.74 | 0.45 | 0.30 | 0.26 |
|  | $\rho$ | 1.06 | 1.35 | 1.45 | 0.94 | 0.63 | 0.53 |

**Fig. 12.** Density-potential evolution of CDM200: $f_{vl}$, $f_{ml}$ and $\rho_l$ are the fraction of space, the fraction of mass and the particle density, respectively, below the level $\phi = \phi_l$ in the gravitational potential. $f_v$ and $f_m$ are the differential fraction of space and of mass per potential interval (in units of $10^6$ km/s). The thin lines in (a), (b) and (c) correspond to $z = 25$, the thick lines to $z = 0$. The dotted line in (a) and (b) correspond to a homogeneous distribution. In (c) the solid lines correspond to the space fraction $f_v$ and the dashes lines to the mass fraction $f_m$.

At $z = 25$, $f_v$, $f_m$ and $\rho$ have almost identical values both in ODR and UDR (and in all simulations) because they characterize the (almost) homogeneous initial matter distribution. At redshift $z = 0$ this situation has changed. In all simulations, the mass fractions have increased in ODR and decreased in UDR, clearest in CDM25 where 90% of the mass have gathered in ODR. In the two small simulations the space fraction occupied by the ODR decreases also. This can be hardly seen in the two big simulations, where $f_v^{ODR}$ is approximately equal to $f_v^{UDR}$ from $z = 25$ to $z = 0$. Due to the more significant change in $f_m$ in comparison to $f_v$, the density of UDR $\rho_{UDR}$ decreases with time (for instance $\rho_{UDR} = \frac{1}{3}\rho_{ODR}$ in CDM200). These issues are in agreement with results of previous investigations (Doroshkevich et al., 1997a) using the correlation function and the core-sampling approach. They provide us with additional quantitative characteristics of the evolution of the models.

Generally speaking, Table 2 shows that during matter clustering only a moderate or negligible change takes place of $f_v^{ODR}/f_v^{UDR}$ while the value $f_m^{ODR}/f_m^{UDR}$ changes much more. Almost the same ratios are obtained if the ODR/UDR are defined with respect to the initial potential distribution. From Fig. 11 one could expect that the highest density peaks were significantly correlated with the potential. But this figure consist only of a small slice and might not be representative enough while Table 2 presents results for four complete simulations.

### 4.3. Potential–density correlations.

Qualitatively, there is no doubt that the (negative) cross–correlation between the potential and the density increases with time because the matter concentration in the high dense clumps is always accompanied by the growth of (negative) potential. However, looking for a quantitative characterization, the correlation between the gravitational potential and density fields turns out to be more complicated.

The simplest characteristic of such a correlation is the correlation coefficient $C(z)$, given by

$$C(z) = \frac{\langle \phi \frac{\delta\rho}{\rho} \rangle}{\sqrt{\langle \phi^2 \rangle \langle \left(\frac{\delta\rho}{\rho}\right)^2 \rangle}}. \tag{7}$$

Here $\langle ... \rangle$ denotes a volume average represented in Fourier space as an average over moments of the spectrum (see, e.g., Demianski & Doroshkevich 1992). The potential $\phi$ and the density excess $\delta\rho/\rho = (\rho - \langle\rho\rangle)/\langle\rho\rangle$ must be taken at the same point. The dispersion of the potential $\langle\phi^2\rangle \equiv \sigma_\phi^2$ was defined in Eq. (6). The other parts of Eq. (7) are defined in the same manner:

$$\left\langle \phi \frac{\delta\rho}{\rho} \right\rangle = -\frac{1}{2\pi^2} \int_0^\infty P(z,k)dk, \tag{8}$$

$$\left\langle \left(\frac{\delta\rho}{\rho}\right)^2 \right\rangle = \frac{1}{2\pi^2} \int_0^\infty P(z,k)k^2 dk. \tag{9}$$

Evidently, the correlation coefficient is independent of the normalisation of the power spectrum, but it is sensitive to the nonlinear evolution of the spectral shape. Here the correlation coefficient $C(z)$ was found using the simulated power spectra $P(z,k)$, compare, e.g., Fig. 7. This procedure is similar to the one used in Sect. 3.3 for the calculations of $\xi(z,q)$.

In Fig. 14 the evolution of the correlation coefficient $C(z)$ with redshift z is shown. At the start of the simulations it has about the value of linear theory, after which the magnitude decreases while in the later evolutionary stages ($z \lesssim 1$) it stays almost constant or slowly increases. This unexpected behaviour of $C(z)$ requires more detailed investigations with a wider sample of simulations.

## 5. Conclusions

Here we have described the evolution of the large scale density perturbations using the characteristics of the potential field. We have also investigated the relation of the density perturbations to the potential. The main results can be summarized as follows:

1. For all simulations the measured values of the typical separation between the zero points of potential, $L_0$, are in agreement with the theoretically predictions. For the largest boxes (CDM200 and BSI200 models) an approximately Poisson distribution has been found for all redshifts (with the exception of the large scale tail in the abundance distribution where the influence of the limited number of harmonics becomes important). The strong dependence of the distribution on the box size and the realized power spectrum became obvious. In case of smaller box sizes the different separations are almost equally distributed. This is more typical for the scale free models rather than for the models with the broad band power spectra.

These results illustrate the properties of the scale $L_0$ as a characteristic of the spatial potential distribution. The very broad Poisson like distribution function found for the separation between zero crossings of the potential demonstrates the complicated spatial structure of the potential distribution where both narrow deep potential wells and very extended regions of positive (or negative) potential exist.

2. Our analysis demonstrates (see Figs. 4, 12, 13 and Table 2) that during the evolution the matter is slowly concentrating within $\Phi^-$ regions which coincide roughly with ODR. The initial spatial potential distribution is mapped into the observed large and superlarge scale structure. Thus the initial potential distribution may be used to predict properties of the large scale matter distribution.

3. The potential distribution forms a smooth and coherent field. On small scales (for scales up to 8 and 60 $h^{-1}$ Mpc for CDM25 and CDM200, respectively) the autocorrelation of the potential is positive. This behaviour agrees with the theoretically prediction (see Fig. 8), but at larger scales the correlation vanishes. This is due to the finite box size which strongly distorts the large scale properties of simulations.

4. A complex evolution has been found for the correlation coefficient of potential and density distribution (Fig.14). While intuitively we expect a monotonous growing correlation due to the successive matter concentration in massive clumps, the dimensionless correlation coefficient firstly decreases to $(70-60)\%$ from its initial amount before it starts to increase slightly.

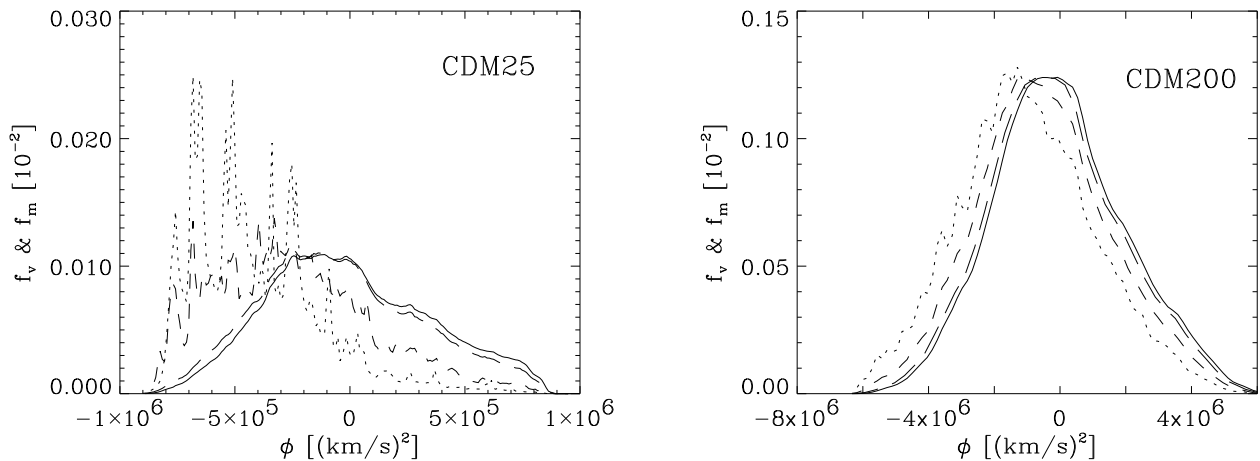

**Fig. 13.** The correlation between the potential at $z = 25$ and the density field at redshifts 25, 2 and 0, for CDM25 & CDM200. Shown are $f_v$ at $z = 25$ (solid line) and $f_m$ at redshifts $z = 25$ (long dashed line), $z = 2$ (dashed line) and $z = 0$ (dotted line)

It is expected that the potential distribution in the universe is traced by the observed galaxy distribution. Indeed, the most promising way for explaining the concentration of galaxies in SLSS elements is to consider the impact of the gravitational potential on structure formation. Also the potential distribution has the required scale range. Thus, we hope to explain the SLSS formation as matter infall to the potential wells, see the more detailed discussion in Demianski & Doroshkevich, 1997. In order to test this hypothesis, we have investigated the main properties of the potential distribution and its time evolution in numerical simulations with different box sizes thereby using power spectra truncated at different scales.

The next step was the comparison of the matter clustering with the potential distribution. As it has been recently shown (Doroshkevich et al. 1997a) for broad band CDM like power spectra there are three evolutionary stages. During the first stage, anisotropic gravitational collapse leads to a fast increase in the number of structure elements with exponentially decreasing mean separation between them. This period is described well by the nonlinear Zeldovich theory. During the second evolutionary stage, the intersection and merging of structure elements dominates, and the slow evolution of the large scale structure is driven by the velocity field. This period can be described by a modified Zeldovich theory tested by Doroshkevich et al. 1997a. During this period, there begins the formation of more massive structure elements. This process dominates the third evolutionary stage when the structure elements disrupt successively to the system of separate massive clumps. Clearly the distinction between these stages is a bit schematic since it depends on the considered scale, and for Gaussian initial conditions the different processes overlap. But it can be expected

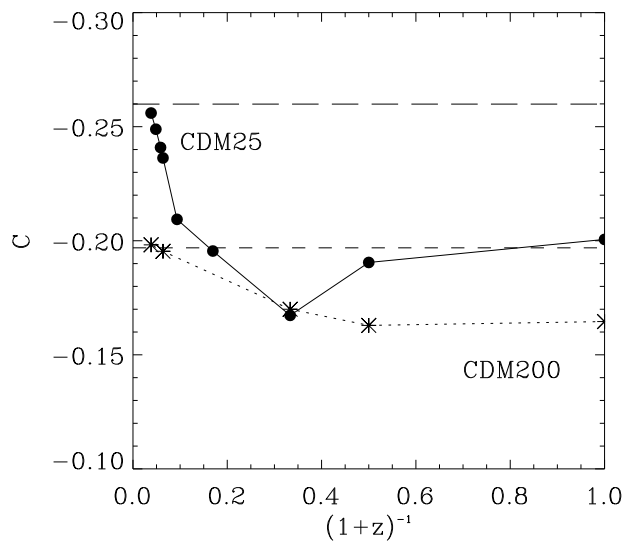

**Fig. 14.** Evolution of the correlation coefficient $C(z)$ with redshift $z$ for CDM25 (full circles) and for CDM200 (stars) vs. redshift $z$. The two horizontal lines show the correlation coefficient calculated by the linear power spectrum (CDM25 long-dashed line, CDM200 dashed line)

that the large–scale potential distribution studied in this paper is responsible mainly for the first stage, and all the detailed relaxation processes are occurring within the initial potential distribution. Let us note however, that for scale free power spectra with negative spectral index typical for the galaxy scale, the second evolutionary stage may become very short, and the relaxation of matter elements in deep potential wells will occur immediately after the pancake formation.

This sequence of processes is obvious in the evolution of the potential distribution. The successive intersection and merging of the structure elements results in the collapse of matter in regions of high potential gradients leading to a small number of narrow deep wells and to a smoothing of the overall potential distribution. The appearance of very massive sheetlike structure elements with a large separation between them has been recently found in one simulation. They remind us in many aspects on the SLSS walls found in observed galaxy distribution, but the simulation cannot reproduce it completely (see Doroshkevich et al. 1997b, and Demianski & Doroshkevich 1997, for a more detailed discussion). Perhaps, the difference is explained by the truncated power spectrum used in simulations. On the other hand it may also result from a large scale bias between the dark matter and the galaxies.

Considering the gravitational potential perturbations we must necessarily improve also the definition of the scale of homogeneity in the matter distribution in the universe. Usually this scale is assumed to be a few times larger than the scale of nonlinearity as defined e.g. by Efstathiou et al. (1988). However, the latter characterizes only the regions of strong matter clustering. Thus, the scale of nonlinearity is similar to the typical scale of the LSS which arises by the merging of structure elements due to their peculiar motion. The spatial distribution of SLSS elements can be understood by the typical scales of the potential distribution. This scale is neither linked directly with matter displacement nor with the scale of nonlinearity. It characterizes the spatial structure of the primordial perturbations.

The gravitational potential perturbations have a strong influence on the matter distribution of the largest scales that leads probably to the appearance of SLSS elements which comprises a significant fraction of all galaxies.

*Acknowledgements.* We are grateful to our referee, Alain Blanchard, for his constructive remarks to the manuscript which helped much in improving the discussion. This paper was supported in part by Danmarks Grundforskningsfond through its support for an establishment of Theoretical Astrophysics Center. This work was also supported by grant INTAS-93-68 and by the Center of Cosmo-Particle Physics, "Cosmion" in the framework of the project "Cosmoparticle Physics".